\pdfoutput=1
\documentclass[sigconf,screen]{acmart}
\usepackage{listings}
\usepackage{algorithm}
\usepackage{algorithmic}
\usepackage{booktabs}
\usepackage{graphicx}
\usepackage{xcolor}

\usepackage{amssymb}
\usepackage{amsmath}
\usepackage{microtype}

\setcopyright{cc}
\setcctype{by}
\acmDOI{10.1145/3832783.3834416}
\acmYear{2026}
\copyrightyear{2026}
\acmISBN{979-8-4007-2882-2/2026/10}
\acmConference[ASE '26]{Proceedings of the 41st IEEE/ACM International Conference on Automated Software Engineering}{October 12--16, 2026}{Munich, Germany}
\acmBooktitle{Proceedings of the 41st IEEE/ACM International Conference on Automated Software Engineering (ASE '26), October 12--16, 2026, Munich, Germany}
\received{2026-03-26}
\received[accepted]{2026-06-18}

\begin{document}

\title{ReqEvolve: User-Oriented Software Self-Evolution through Automatic Requirement Interpretation}

\author{Md Asif Iqbal Fahim}
\orcid{0000-0001-9814-9943}
\affiliation{%
  \institution{University College Dublin}
  \department{School of Computer Science}
  \city{Dublin}
  \country{Ireland}
}
\email{md.fahim@ucdconnect.ie}

\author{Alessio Ferrari}
\orcid{0000-0002-0636-5663}
\affiliation{%
  \institution{University College Dublin}
  \department{School of Computer Science}
  \city{Dublin}
  \country{Ireland}
}
\affiliation{%
  \institution{Consiglio Nazionale delle Ricerche (CNR)}
  \city{Pisa}
  \country{Italy}
}
\email{alessio.ferrari@ucd.ie}

\begin{abstract}
The paradigm of software self-evolution enables systems to autonomously extend and reconfigure their own capabilities during execution in response to technical specifications. Yet requests for new functionality often originate from end users and are rarely expressed in technical terms. As a result, developers must translate user needs into technical specifications before the system can evolve, delaying early validation of the requested functionality by preventing users from immediately observing the resulting behaviour. To address this gap, we present \textsc{ReqEvolve}, a runtime code generation system that enables \textit{user-driven} self-evolution by accepting high-level user requests. The system integrates automatic requirements engineering (RE) and test-driven development (TDD) to transform these requests into executable functionality through clarification, specification decomposition, test generation, and runtime integration. We evaluate \textsc{ReqEvolve} on 72 software evolution cases across 18 projects against two baselines: SpecFix, an RE-focused code generation approach, and an ablation variant of our system.
\textsc{ReqEvolve} achieves 89.2\% Pass@1, outperforming SpecFix by 18.8\% ($p < 0.01$, $r = 0.79$, large effect) and the ablation baseline by 32.6\% ($p < 0.001$, $r = 0.88$, large effect).
These results provide initial evidence that user-driven self-evolution is a viable paradigm for autonomously extending software capabilities from user requests, thereby accelerating requirements validation prior to developer verification.
\end{abstract}



\maketitle

\section{Introduction}
Software systems operate with fixed capabilities determined at development time, making functional changes costly and slow~\cite{dora2024}. Recent LLM-based code generation tools accelerate software development by approximately 21\%~\cite{paradis2025much}, with commercial systems such as GitHub Copilot~\cite{copilot} and research approaches such as CodeT5~\cite{wang2021codet5}, MetaGPT~\cite{hong2023metagpt}, and AgentCoder~\cite{huang2024agentcoder} supporting automated code synthesis. However, these approaches primarily target static code generation or inline completion from implementation-oriented requests. This is limiting in \textit{exploratory} development, where requirements are often underspecified and stakeholders are more interested in rapidly assessing the behaviour of a new capability than in first obtaining and inspecting its static implementation.

To address this limitation, the paradigm of \textit{software self-evolution} has emerged~\cite{fahim2026selfevolve}. In self-evolution, a developer requests that the software implement a novel functionality, potentially requiring architectural restructuring, and the system autonomously synthesises, integrates, and executes the new capability at runtime. Rather than first inspecting generated source code, the developer can immediately observe the resulting behaviour and assess whether the requested capability is useful and aligned with the intended goal.

Current self-evolution approaches remain developer-mediated. Nevertheless, the intent motivating a new capability often originates at the \textit{user} level and is expressed in domain terms
(e.g., ``I want to discover people to connect with'').
Existing approaches do not consume such intent directly, but instead assume that it has already been reformulated into an implementation-oriented specification
(e.g., ``create a friend\_suggestions function that recommends friends from social\_graph.json in PROJECT KNOWLEDGE with progressive enhancement based on RECOMMENDATION\_MODE environment variable. [...]''---from the self-evolution proposal by Fahim et al.~\cite{fahim2026selfevolve}).
Consequently, translating user-expressed needs into technical specifications remains a bottleneck in the development process, besides being a major source of ambiguity~\cite{ferrari2015ambiguity}.

To address this gap, we propose the concept of \textit{user-driven} self-evolution, i.e., the ability for software to generate and integrate new functionality during execution based on end-user requests. Instead of waiting for developers to implement a requested feature, users could express their needs through a natural-language interface embedded in the software itself and immediately observe working behaviour, thereby enabling early \textit{validation} of requirements, i.e., checking whether the ``right software'' is being built. This can collapse feedback from weeks to minutes, fostering faster product improvement before developers \textit{verify} correctness and non-functional properties, i.e., whether the software has been ``built right''.

To support this novel paradigm, we propose \textsc{ReqEvolve}, an agentic AI platform including an automatic requirement engineering (RE) pipeline, combined with  test-driven development (TDD). The former takes a user request and performs user need clarification and specification refinement. The latter enables test generation followed by functionality synthesis and runtime integration.
We evaluate  \textsc{ReqEvolve} on a novel benchmark of self-evolution scenarios, each involving integration of new capabilities into existing codebases (40-3,100 LOC). We compare our solution with a compatible, yet static, code generation tool, namely SpecFix~\cite{specfix2025}, and we perform an ablation study. \textsc{ReqEvolve} achieves 89.2\% Pass@1, outperforming SpecFix by 18.8\% ($p < 0.01$, $r = 0.79$) and the ablation baseline by 32.6\% ($p < 0.001$, $r = 0.88$). We also perform a root cause analysis to identify the sources of failures, showing that most of the cases (64.5\%) are due to the code generation agent, which hallucinates even in presence of clear input and correct requirements refinement.
The contribution of this paper is three-fold: (1) we introduce the novel concept of user-driven self-evolution; (2) we provide an implementation of the paradigm, named \textsc{ReqEvolve}; and (3) we develop a benchmark of 72 evolution cases across 18 projects, the first one specifically targeting the novel concept.

\section{User-Driven Self-Evolution}

\begin{figure}[t]
\centering
\includegraphics[width=\columnwidth]{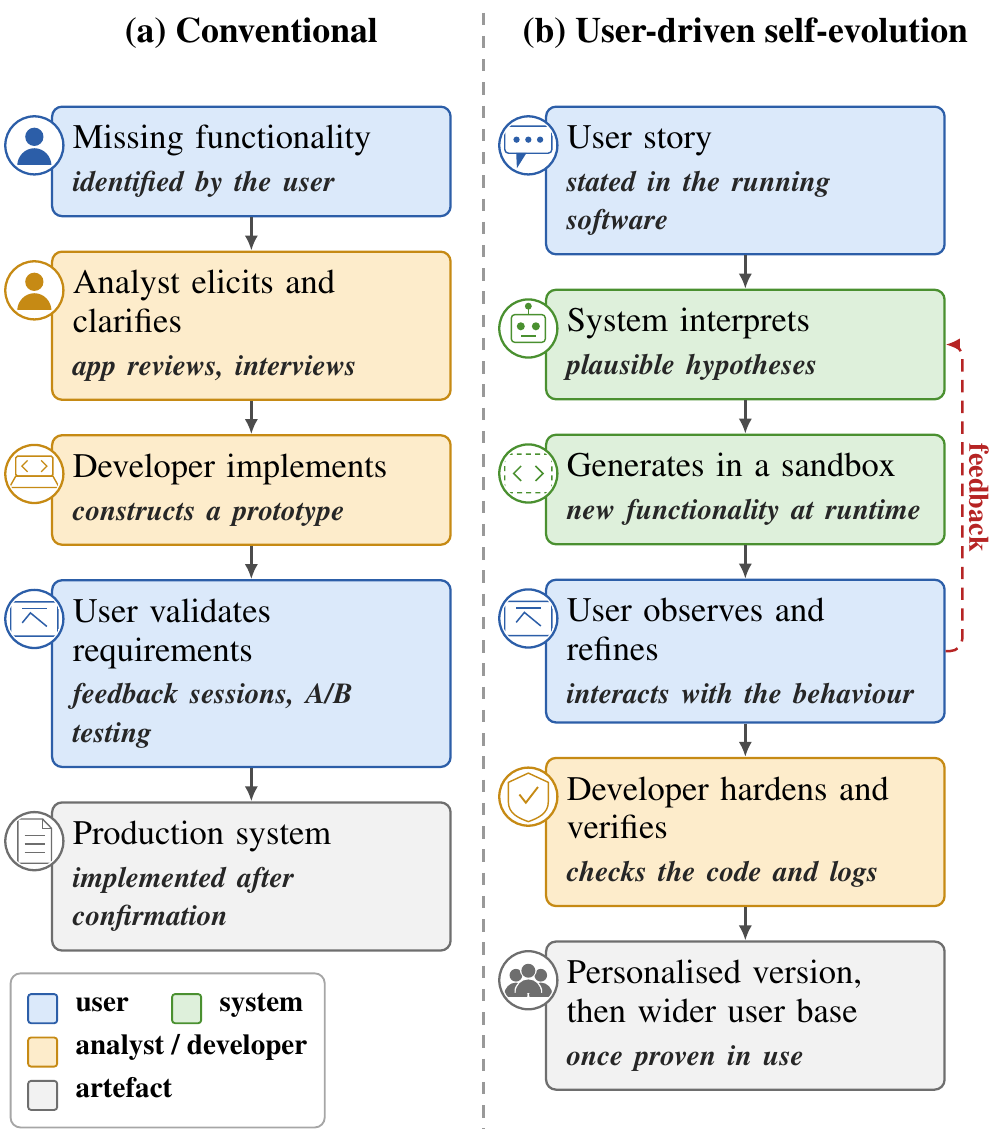}
\caption{Conventional software evolution (a) versus user-driven self-evolution (b).}
\label{fig:user-driven}
\end{figure}

This section introduces the concept of user-driven self-evolution, and contrasts it with a more conventional software evolution approach.
Both paradigms begin from a software system with an existing set of capabilities. The difference lies in how missing functionality is elicited, translated, and realised. Figure~\ref{fig:user-driven} contrasts the two processes.

In a conventional process, when users identify a missing functionality, this need is elicited  by a requirements analyst, for example through app reviews, interviews, or other feedback channels. The analyst then refines and, in case of interviews, clarifies the requirement before handing it over to developers for implementation. In an LLM-based variant of this process, developers may use generative tools to explore design alternatives and construct a prototype more quickly.
The prototype is then returned to users for requirements validation, for example through feedback sessions or A/B testing, and only after confirmation is the functionality fully implemented and integrated into the production system.

By contrast, in our vision of user-driven self-evolution, the user expresses a need directly through a natural-language interface, using a user story as the interaction format. We adopt user stories because they are an established RE artefact supported by validated quality assessment frameworks~\cite{lucassen2016aqusa}. Although user stories are not the most natural way for users to interact with software applications, they can serve as a structured form of human-machine communication. For instance, a user might state: ``As a user, I want the system to suggest relevant people to connect with, so that I can build my professional network around shared interests.'' Since user requests are often underspecified, the system formulates and evaluates plausible hypotheses of interpretation. In the previous example, these hypotheses might include suggesting nearby users with similar interests, recommending collaborators based on past interactions, or highlighting users active in the same discussion spaces. Based on the most plausible interpretation---selected based on internal reasoning, user context, and knowledge of the current software and its domain---the system generates the functionality in a sandbox and allows the user to immediately observe the output and interact with the resulting behaviour. The user can then provide further feedback, iteratively refining the capability until it is considered satisfactory. At that stage, the generated functionality may still exhibit security weaknesses or fail to meet non-functional requirements (performance, security, regulatory). Furthermore, seemingly correct outputs may still hide implementation defects. Developers check the code and logs produced by the system, and then harden, verify, and update the feature before broader deployment. Initially, this yields a personalised version of the software for the requesting user, while ``proven in use'' capabilities, which have gone through further validation, can subsequently be generalised and released to the wider user base. At this stage, we use \textit{self-evolution} to denote \textit{self-extension}, i.e., the autonomous addition of new functionality, rather than self-refactoring or broader forms of structural adaptation, although the underlying development model could accommodate such extensions in future work.

\section{Background and Related Work}\label{sec:background}

\subsection{Background}

\textbf{User Stories in Agile Development.} User stories are a lightweight requirements artifact in agile software development~\cite{cohn2004userstori}. A user story captures a capability request using the format: ``As a [role], I want [means], so that [ends].'' The three components, role (who makes the request), means (what capability they want), and ends (why they need it), express user needs without prescribing implementation. For instance, ``As a data analyst, I want to calculate average salaries, so that I can analyze compensation trends'' specifies WHAT (calculate averages) and WHY (analyze trends) without dictating HOW (algorithm, data structure, or implementation approach).
User stories are designed to express \textit{functional} requirements. \textit{Non-functional} requirements are typically more technical, and expressed as ``shall'' requirements (e.g., ``The system shall scale to support 10{,}000 concurrent users'').

\textbf{User Story Quality with AQUSA.} The AQUSA (Automatic Quality User Story Artisan) framework~\cite{lucassen2016aqusa} assesses user story quality based on the 13 quality criteria, addressing syntactic, semantic, and pragmatic defects. AQUSA checks for well-formed role-means-ends structure and detects defect categories, including missing components (role, means, or ends), conjunctions bundling multiple features into a single story, and inconsistent formatting.

\textbf{Acceptance Criteria.} Acceptance criteria translate user stories into testable specifications using the GIVEN-WHEN-THEN format~\cite{north2006bdd}: ``GIVEN a context, WHEN an action occurs, THEN an outcome results.'' Acceptance criteria may comprise multi-step sequences describing complex workflows, e.g., ``GIVEN employee records and department data, WHEN the analysis function is invoked, THEN it retrieves records, groups by department, computes averages, and returns a sorted summary.'' Acceptance criteria bridge user-expressed needs (i.e., user stories) and implementable specifications by defining success conditions without dictating implementation mechanisms.
We are not aware of established quality frameworks or tools specifically targeting acceptance criteria.

\subsection{Related Work}
Our contribution intersects multiple research areas: LLM-based code generation, including RE-focused approaches, self-evolution, and code generation benchmarks. We position our contribution relative to each area.

\textbf{LLM-based Code Generation.} LLM-based code generation approaches employ agent-based architectures, with some using single-agent systems and others coordinating multiple agents.

\textit{Single-agent} approaches evolved progressively to address performance and architectural gaps. Pre-trained encoder-decoder models (e.g., CodeT5~\cite{wang2021codet5}) established code-specific architectures, while planning-based approaches~\cite{jiang2024self} decompose coding tasks into numbered steps before implementation.
However, unvalidated plans create a failure risk. Self-Debugging~\cite{chen2023selfdebug} addresses this limitation through execution-based feedback loops, where models iteratively debug their own code through test execution and ``rubber duck'' explanation.
This purely reactive approach fixes only after failure with no persistent learning. Reflexion~\cite{shinn2023reflexion} addresses this by adding episodic memory for verbal self-reflection across problems
AlphaCodium~\cite{ridnik2024alphacodium} unified these insights through test-driven flow engineering with progressively harder test execution.
Commercial tools such as GitHub Copilot~\cite{copilot} demonstrate widespread adoption in development environments.

\textit{Multi-agent} frameworks extend single-agent approaches by coordinating specialised roles for complex synthesis. MetaGPT~\cite{hong2023metagpt} implements workflows as a 5-role pipeline (Product Manager, Architect, Engineer, QA) communicating via structured documents.
AgentCoder~\cite{huang2024agentcoder} simplifies to 3 agents (Programmer, Test Designer, Test Executor), where the critical innovation is Test Designer generating tests independently from code, preventing confirmation bias
ChatDev~\cite{chatdev} employs 7+ role-playing agents in chat chains for collaborative software development. AutoGen~\cite{wu2023autogen} provides a general multi-agent conversation framework yet acknowledges that tasks without back-and-forth troubleshooting may not benefit from coordination overhead. Recent research ~\cite{li2026single} shows compiling multi-agent systems into focused mechanisms reduces latency,
suggesting that functional benefits can be achieved without multi-agent coordination costs.

\textbf{RE-focused Code Generation.} This group of works encompasses contributions on code generation that focus on the RE dimension, considering requirements not as a mere input, but as something to be clarified and manipulated before being fed to a code generation pipeline.
According to the systematic literature review by Hemmat et al.~\cite{hemmat2025relr}, most LLM-for-RE studies focus on isolated RE tasks,  and do not evaluate their effect on downstream code generation. As a result, RE-focused code-generation approaches remain limited. Among them,
ClarifyGPT~\cite{mu2024clarifygpt} detects ambiguity in programming problem descriptions and generates clarifying questions for users before code generation. SpecFix~\cite{specfix2025}, the current state-of-the-art, automates specification repair, outperforming ClarifyGPT and other approaches across multiple benchmarks. These systems work on technical specifications provided by developers.

\textbf{Runtime Self-Evolution.} Self-adaptive systems reconfigure existing components at runtime via MAPE-K (Monitor-Analyze-Plan-Execute-Knowledge) control loops~\cite{mape-k,seams2020}, the dominant reference architecture for self-managing systems. Classical MAPE-K operates on pre-defined adaptation strategies, parameter tuning, component reconfiguration, and resource scaling, but cannot generate new code. Li et al.~\cite{li2024genaisas} showed how LLMs could serve as reasoning components for each MAPE-K phase, yet this remained largely theoretical with few production implementations. SelfEvolve~\cite{fahim2026selfevolve}, the only existing framework capable of runtime self-extension, demonstrated actual runtime code generation through test-driven development (TDD), outperforming development-time multi-agent frameworks.

\textbf{Code Generation Benchmarks.} Code generation benchmarks have progressed toward realism yet reveal evaluation gaps. HumanEval ~\cite{humaneval} (164 function-level problems) and MBPP~\cite{mbpp} (974 problems) are among the most widely used code generation benchmarks. Both mainly evaluate synthesis of small, self-contained Python solutions from natural-language problem descriptions: HumanEval focuses on generating function implementations from docstrings, while MBPP contains short Python programming tasks intended for entry-level programmers.
CodeContests~\cite{codecontests} (4,040 competitive programming problems) tests algorithmic reasoning, but competitive programming differs fundamentally from real-world software engineering. ClassEval~\cite{du2024evaluating} focuses on object oriented programming, revealing a dramatic performance drop: GPT-4 falls from 85.4\% on HumanEval to 37.0\% on class-level tasks, proving function-level ability does not transfer to class-level. CoderEval~\cite{yu2024codereval} (460 tasks from 43 real projects) introduces context dependency levels, finding >70\% of real-world functions are non-standalone, yet still evaluates against fixed specifications extracted from existing code.

\textbf{Our Contribution.} \textsc{ReqEvolve} is the first system enabling runtime code generation from user stories. All surveyed approaches (ClarifyGPT, SpecFix, code generation systems, self-evolution systems) operate on developer-oriented inputs: technical specifications, programming problem descriptions, or formal specifications. We uniquely accept user stories, the agile standard for capturing end-user requirements.  Compared to development-time tools, we target \textit{runtime} capability extension, similar to the preliminary short contribution from Fahim et al., i.e., SelfEvolve~\cite{fahim2026selfevolve}.
In our work, we incorporate the SelfEvolve TDD pipeline, as it is the only solution supporting self-evolution.

Another relevant contribution is our benchmark of 72 evolution cases across 18 projects. Unlike existing benchmarks, ours targets the novel concept of user-driven self-evolution. Although prior benchmarks contain more problems (HumanEval: 164, MBPP: 974, ClassEval: 100, CoderEval: 460, CodeContests: 165), our cases are substantially more complex. Each case requires \textit{runtime integration} of functions of ${\sim}$62 LOC with an existing codebase of 40--3,100 LOC, yielding a more challenging setting than the mostly self-contained functions in HumanEval (${\sim}$11 LOC), MBPP (${\sim}$7 LOC), CodeContests (${\sim}$60 LOC), ClassEval (${\sim}$46 LOC), and CoderEval (${\sim}$30 LOC). In contrast to these datasets, our benchmark captures realistic generation scenarios that require compositional reuse, and data-processing tasks involving external data.
While 11 of the 18 projects reuse and substantially extend codebases from the initial SelfEvolve package (which was limited to projects below 800 LOC), the associated 72 evolution cases are entirely new, being defined by a novel set of user stories and requested capabilities rather than by the original SelfEvolve tasks.

\section{\textsc{ReqEvolve} Architecture}\label{sec:architecture}

\textsc{ReqEvolve} implements a layered pipeline architecture for runtime code generation from user stories. The system comprises multiple components. Most are \textit{agents}, i.e., LLM instances configured with component-specific prompts and tool access. A few components operate without LLM invocation, providing rule-based validation, test execution, audit logging, or persistent storage. The nature of the components is specified in their detailed description in the following paragraphs.

At a high level, the pipeline executes two main stages: an \textit{RE stage} that clarifies user intent and infers missing information, rephrases the request, and decomposes it into acceptance criteria; a \textit{test-driven code generation stage} that iteratively produces and validates implementations, following the SelfEvolve idea. These two processing stages are supported by a \textit{Persistent Knowledge Base}, a data infrastructure layer maintaining pre-existing and generated functions, prompt templates for the agents, and tool descriptors for compositional reuse and immediate function discovery. A distinguishing capability: generated functions can utilise functions from the Knowledge Base as sub-routines, enabling compositional building.

\begin{figure}[t]
\centering
\includegraphics[width=0.8\columnwidth]{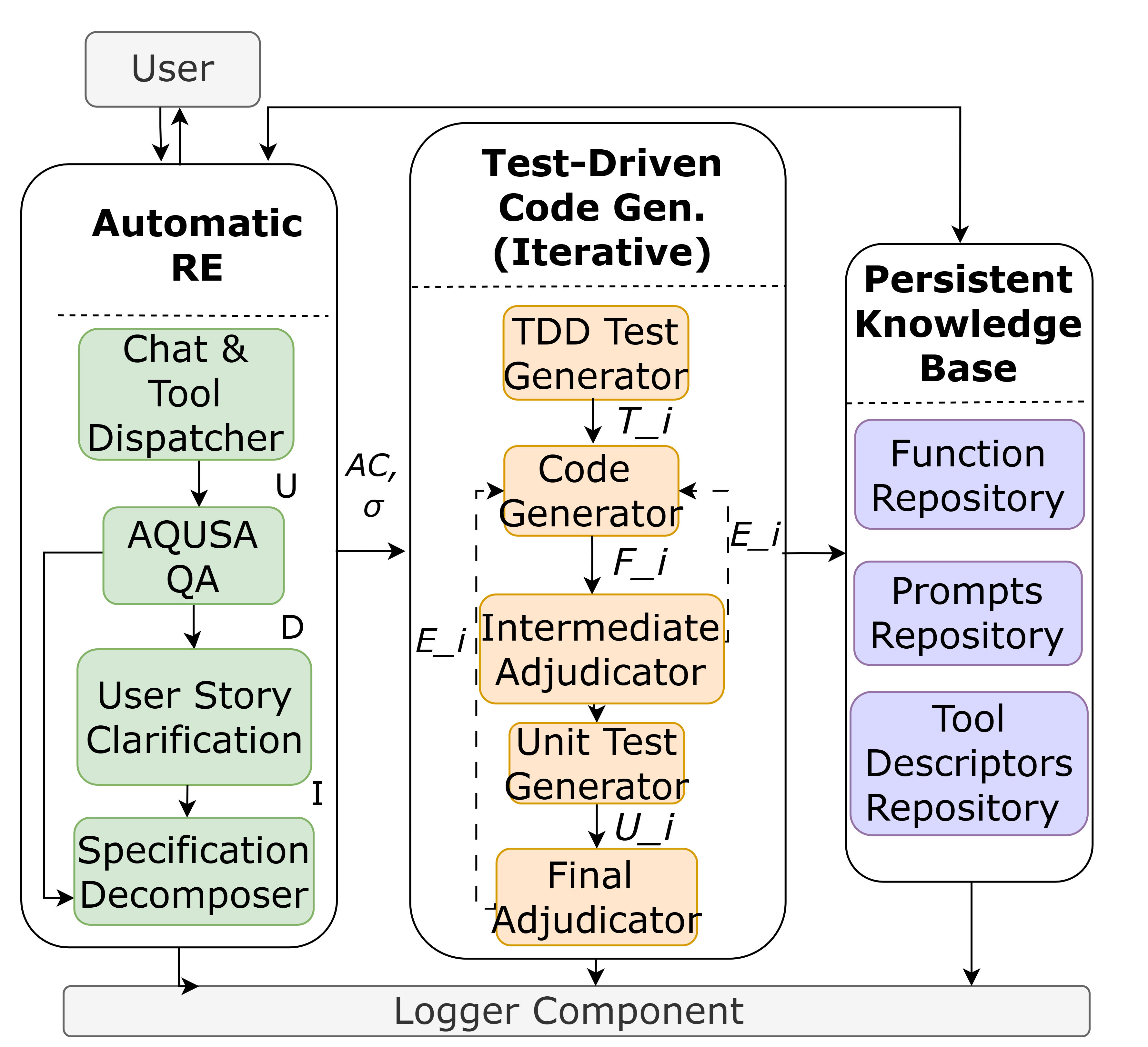}
\caption{\textsc{ReqEvolve} architecture.
}
\label{fig:architecture}
\end{figure}

The detailed architecture is depicted in Figure~\ref{fig:architecture}. The pipeline proceeds through the two stages described above in an incremental and iterative manner.
Each iteration maintains the current state to enable refinement based on test feedback. More formally, the pipeline state at iteration $i$ consists of: user request $U$, codebase context $C$ (relevant code from the project), defect set $D$ from a quality assessment component, requirements interpretation hypotheses $I$, acceptance criteria $AC = \{AC_1, \ldots, AC_m\}$, function signature $\sigma$, TDD tests $T_i$, function code $F_i$, and feedback $E_i$. We use $\text{agent}(\cdot)$ to denote LLM invocation with component-specific prompts from the Prompts repository.

We illustrate the pipeline operation using a social network application as our running example. Consider a social network application with a \texttt{UserNetwork} class providing relationship queries, a database of user profiles and connections, and recommendation utilities.
A user submits a request through a partially complete user story: \textit{``I want to discover people to connect with''}.

\subsection{System Components}

\subsubsection{Persistent Knowledge Base.}

The Knowledge Base comprises three repositories. The \textit{Prompts} repository stores system instructions for each LLM-based  component. The prompts specify the component's task, input format, and expected output structure. These were iteratively developed following OpenAI best practices\footnote{\url{https://developers.openai.com/api/docs/guides/prompt-engineering}}.
The \textit{Functions} repository contains executable Python implementations of generated functions (e.g., to show a list of connections, or the dashboard). The \textit{Function Descriptors} repository maintains JSON specifications (function name, parameters, descriptions, invocation metadata) pairing functions with their metadata, enabling function-calling (where an LLM selects and invokes registered functions).

\subsubsection{Automatic Requirements Engineering.}

This layer comprises three agents and one rule-based component: Chat \& Tool Dispatcher (discovers existing functions and prepares codebase context); AQUSA Quality Assurance  (assesses user story quality and detects defects, rule-based), Clarification (resolves defects and adds behavioral context), and Specification Decomposer (translates clarified stories into acceptance criteria and function signatures).

\textbf{Chat \& Tool Dispatcher.}\label{sec:dispatcher} The entry component receives the request $U$ and searches available Function Descriptors from the Knowledge Base.
If an existing function matches semantically, the agent returns a tool call \texttt{tool\_call}$(f_j)$, the function executes immediately, and the results are returned to the user without generation. Generation occurs only for requests that do not match any existing function, preventing redundant implementation.

When no existing function matches, the dispatcher prepares the codebase context $C$ for generation. Loading the entire project codebases would exceed the token limits of the LLMs (typical projects contain thousands of lines across dozens of files). The dispatcher employs selective context retrieval using automated code parsing and metadata extraction: extracting lightweight metadata, invoking the agent to select relevant components semantically, and retrieving complete implementations for only those selected items (classes with their methods, data file schemas). This reduces context
to a manageable size while preserving necessary dependencies. The dispatcher forwards $U$ to the AQUSA Quality Assurance (QA) component for validation. The curated context $C$ is retained for use by downstream code generation components.

\textbf{AQUSA QA.} The component receives user request $U$ from the dispatcher and validates it using the AQUSA framework~\cite{lucassen2016aqusa}.
AQUSA QA outputs the defect set $D$ (empty if well-formed). For our example, $U = $ ``I want to discover people to connect with,'' AQUSA detects $D = \{\text{no\_role}, \text{no\_ends}\}$ (missing stakeholder role and missing benefit). The system passes $D$ to the clarification component.

\textbf{User Story Clarification.} This component aims to clarify the user story based on the detected defects. Rather than querying the user, the agent autonomously hypothesises plausible clarification answers, enabling generation without user interaction. The assumption is that it is easier for users to give feedback on a prototypical implementation, rather than through questions.
More formally, when AQUSA detects defects ($D \neq \emptyset$), the agent generates clarification hypotheses using the user story $U$, defect set $D$, and codebase context $C$ to improve the user story.

Hypotheses target four categories. The first three categories, namely \textit{Stakeholder, Scope, Outcome}, are associated with the fields of the user story (role, means, end). The hypotheses aim to identify the missing information in that field, or to clarify its meaning. The last category, \textit{Example}, provides usage scenarios. More specifically:
\textit{Stakeholder} resolves missing or vague roles, grounding the ``As a [stakeholder]'' component. \textit{Scope} establishes boundaries and constraints, refining ``I want [action]'' into a specific, bounded request. \textit{Outcome} clarifies expected behavior and edge cases, completing ``so that [benefit]'' and defining success conditions. \textit{Example} provides concrete usage scenarios that helps the other pipeline components contextualise the user story.

The agent generates clarification  hypotheses, formed by clarification questions and candidate answers. Answers are ranked by the agent according to contextual plausibility. Although this ranking is heuristic, it is not critical, as users can refine the functionality generated based on the hypotheses after observing its behaviour.
The clarification questions and top-ranked answers form $I$, the requirements interpretation hypotheses set.
For our example with $D = \{\text{no\_role}, \text{no\_ends}\}$ (missing stakeholder and missing benefit), the following interpretation hypotheses are generated: (1) ``Who will use this feature?'' $\rightarrow$ ``Social network user'' (addresses no\_role). (2) ``What benefit do you seek?'' $\rightarrow$ ``Expand my social network'' (addresses no\_ends). The requirements interpretation hypotheses $I$ guides the subsequent specification decomposition.

It should be noted that our pipeline performs capability matching only once, on the initial user request. As a result, if later clarification resolves defects in the user story, it is possible that the refined request would in fact correspond to an already existing function. We currently do not repeat matching after clarification, as once a request is deemed novel, later iterations focus on refining the new capability rather than revisiting reuse. This simplifies the control flow, but may lead to missed reuse opportunities and therefore represents a limitation of the current approach. Also, at this stage, we discard other candidate answers, but these can, in principle, be reused in further iterations in case the hypotheses result in being incorrect, after user validation. It should also be noted that our framework always attempts to find an answer, regardless of the actual quality of the user story---i.e., no input is rejected. In a real-world scenario, this behaviour would likely need to be complemented by input validation and safeguard mechanisms to detect requests that are too incomplete or ambiguous.

\textbf{Specification Decomposer.} The decomposer receives the original user request $U$ and requirements interpretation hypotheses $I$ from the previous step. It invokes the agent at low temperature for consistency to perform two operations. First, it constructs a proper user story from the incomplete input by incorporating the $I$ set, transforming ``I want to discover people to connect with'' into a complete user story following AQUSA format: ``As a social network user, I want to discover potential friend connections from my network, so that I can expand my social circle with relevant people'' (role: ``social network user,'' means: ``discover friend connections,'' ends: ``expand social circle''). Second, it translates this generated story into acceptance criteria $AC$ and function signature $\sigma$. For our example, $ACs$ are: GIVEN a user with network connections, WHEN requesting friend suggestions THEN it returns a list of suggested profiles; GIVEN a user with no connections, WHEN the discovery executes THEN it returns an empty list. The inferred function signature is \texttt{def discover\_friends(user\_id: str) -> list}; the implementation is generated by the next component.

\subsubsection{Test-Driven Code Generation.}
This layer comprises four agents and one execution component: TDD Test Generator and Code Generator (generate tests from acceptance criteria, and implementation code, respectively), Intermediate Adjudicator (verifies implementation against TDD tests), Unit Test Generator (creates comprehensive test suite), Unit Test Handler (executes tests in isolated subprocesses and captures results, without LLM invocation), and Final Adjudicator (evaluates unit tests and integrates verified functions).

\textbf{TDD Test Generator and Code Generator.} With acceptance criteria from the decomposer, the pipeline enters the iterative code generation loop, which is based on the architecture from Fahim et al.~\cite{fahim2026selfevolve}. This component has two responsibilities that execute sequentially. First, it generates TDD test cases:

$T_i = \text{agent}(U, AC, \sigma, E_{i-1})$ where $E_{i-1}$ contains feedback from the previous iteration (empty initially). Tests derive directly from acceptance criteria. Each GIVEN-WHEN-THEN statement becomes a test function using real codebase objects from the retrieved context, not mocks. The signature $\sigma$ enforces strict naming so test imports match implementation exactly.

Second, it generates function implementation:  \-

$F_i = \text{agent}(T_i, AC, \sigma, E_{i-1})$. The generator examines available codebase classes and previously generated functions from the Functions repository, importing and reusing them rather than reimplementing. The generator can also import standard libraries and external packages (NumPy, Pandas, scikit-learn) from the Python environment when the task requires specialized computation. For our example at iteration 1 ($E_0 = \emptyset$), the generator produces test functions verifying that the result is a list of user profiles, that users with no mutual connections return an empty list, and that profiles with missing data are filtered. It then generates \texttt{discover\_friends(user\_id)}, which retrieves the user's network, identifies mutual connections, ranks candidates, filters incomplete profiles, and returns a list of suggested friend profiles.

\textbf{Intermediate Adjudicator.} The pipeline employs two-phase adjudication to separate concerns: intermediate verification checks whether the implementation satisfies acceptance criteria, while final verification probes implementation robustness. This separation prevents wasting resources on comprehensive testing of semantically incorrect code. The intermediate adjudicator verifies the generated function against the TDD tests written alongside it. The adjudicator executes: $\text{test\_output} = \text{execute\_tests}(F_i, T_i)$, capturing stdout, stderr, and exit codes with a timeout. It then invokes the agent on the test output: $\text{agent}(\text{test\_output}) \rightarrow \{\textit{accept}, \textit{reject}\}$. On \textit{accept}, the pipeline advances to unit test generation. On \textit{reject}, the adjudicator produces feedback $E_i$ specifying the failure phenomena import errors, wrong return types, incorrect numerical results, missing edge case handling and control returns to the TDD Test Generator and Code Generator for iteration $i+1$. For our example, suppose iteration 1 generates a function that returns random users rather than mutual connections: the test for connection-based suggestions fails, the adjudicator rejects with feedback ``Function returned users without mutual connections; user specified `close connections (friends of friends)' but implementation uses random sampling,'' and iteration 2 generates corrected code prioritizing mutual connections as specified, passing all TDD tests.

\textbf{Unit Test Generator.} After intermediate acceptance, the pipeline generates a separate, more comprehensive test suite examining the specific implementation of $F_i$ rather than just high-level requirements. Unlike TDD tests (which encode acceptance criteria), unit tests probe implementation robustness: $U_i = \text{agent}(F_i, R)$. These tests cover edge cases specific to how the function was actually written, not just what it should do, but how it behaves under stress. For our example, the unit test generator produces tests for: a user with no connections (should return an empty list), a user with a single mutual friend (returns that one suggestion), and a user with incomplete profile data in their network (should filter out invalid profiles and return only complete suggestions).

\textbf{Unit Test Handler.}     The handler executes the unit test suite $U_i$ through an isolated subprocess with a timeout, capturing full execution state: stdout, stderr, test counts (pass and fail counts), and exit codes. Timeout enforcement prevents infinite loops in generated code from blocking the pipeline. The handler returns a structured result record to the Final Adjudicator for assessment.

\textbf{Final Adjudicator and System Integration.} The second adjudicator evaluates unit test results via $\text{agent}(\text{unit\_test\_output}) \rightarrow \{\textit{accept}, \textit{reject}\}$. On \textit{accept}, the function is promoted: its code is written to the Functions repository, its JSON tool descriptor is registered in the Function Descriptors repository, enabling future discovery by the Chat \& Tool Dispatcher, and the Prompts repository is updated with patterns for when to invoke this function, constituting a permanent capability addition without restart. On \textit{reject}, the adjudicator produces dual feedback: a code correction suggestion targeting the specific implementation failure, and a unit test regeneration trigger (when failures repeat without progress) to prevent test staleness. After $n$ consecutive failures with identical test errors, the system forces full test regeneration from scratch. Control returns to the Test-Driven Code Generator for iteration $i+1$. The combined iteration count across both adjudicators is bounded at a configurable maximum $K$ to prevent infinite loops. For our example, iteration 2's corrected function passes both TDD tests and unit tests, completing in 2 iterations total with \texttt{discover\_friends(user\_id: str) -> list} integrated into the Knowledge Base and immediately available for future requests.

\subsection{Logger Component}

The system logs all pipeline stages to JSON files with a consistent schema. Each log entry contains: stage identifier (e.g., ``aqusa\_qa'', ``clarification''), timestamp (ISO 8601), iteration number, input data, output data, and execution metadata. The logs capture outputs from the RE pipeline (e.g., AQUSA defects, decomposed requirements, acceptance criteria), from TDD iterations (test code, implementation code, adjudicator verdicts with rejection reasons), validation results (test outcomes, failure messages), and metadata (timestamps, iteration numbers, token counts).

\textbf{Developer-oriented Design.}  We adopt three pragmatic design principles, motivated by requirements traceability~\cite{gotel1994traceability,vogelsang2025impact} and dependability/failure-analysis~\cite{Avizienis2004} concepts: \textit{semantic traceability}: each log entry links user intent to implementation decisions e.g., when clarification infers ``close connections,'' logs record the question, answer, ranking, and resulting requirement;  \textit{failure diagnostics}: logs include failing assertions like \texttt{assert len(results) > 0}, actual outputs, and explanations like ``user story specified `suggest friends' but no connections retrieved''; and \textit{completeness}: all LLM prompt-response pairs logged to verify the absence of hallucination.

 These principles are intended to support three audit tasks developers may perform before deployment: (1) \textit{Specification validation}: verify that the decomposed specifications reasonably match the user story, (2) \textit{Implementation inspection}: examine code, tests, and iteration history for bugs or security issues, (3) \textit{Pipeline performance assessment}: check token and iteration counts for optimization needs. The JSON format supports automated analysis (scripts, dashboards).

This logging enables a two-tier deployment model.

\textbf{Tier 1: User experimental use.} Generated features are immediately available to the requesting user, who validates whether they match intent through direct use (e.g., checking if the friend  suggestions align with their social circle).

\textbf{Tier 2: Developer-audited deployment.} Before propagating features to other users, developers inspect logs to certify quality using the three audit tasks above. For example: a developer opens the logs for a ``friend suggester'' feature, verifies that clarification inferred ``mutual connections'' from the incomplete input, examines the generated code for edge case handling, and checks that generation completed in 3 iterations with acceptable token usage. Afterwards, the features can be propagated to production users.

In our work, we do not evaluate the logger component in isolation, and instead use its outputs only to support failure analysis (cf. RQ3). This limits the strength of our conclusions, since logging inaccuracies could affect the interpretation of failures. However, the logger is an auxiliary, comparatively simple component rather than a core reasoning module, and it was tested conventionally during development to ensure correct recording of extensive execution information. We arguably consider its outputs sufficiently reliable for supporting failure analysis. Whether these logs are sufficient for developer audit in practice remains to be validated empirically and is left for future work; at this stage, we can only claim their utility for the RQ3 analysis.

\section{Evaluation}

Our evaluation aims to answer three research questions (RQs):

\begin{itemize}
    \item \textbf{RQ1:} \textit{What is the performance of \textsc{ReqEvolve}?} We evaluate our solution through a benchmark of 72 software evolution cases across 18 projects, and compare the results with SpecFix, the state-of-the-art RE-focused code generation baseline.
    \item \textbf{RQ2:} \textit{What is the impact of the RE components on generation performance?} We remove the RE components from our pipeline,  feed the requirements directly to the TDD component, and observe the impact on performance.
    \item \textbf{RQ3:} \textit{What are the typical failures and failure patterns?} We perform a thematic analysis of the logs of the failure cases to identify errors, faults, failures, and chains thereof, with the goal of identifying improvement opportunities.
\end{itemize}

\subsection{Implementation and Experimental Setup}

\textbf{Models and Parameters.} All pipeline components employ GPT-4.1 (Azure OpenAI deployment). To ensure fair comparison, the same model is used also for the baselines.
We employ OpenAI's function-calling protocol for function discovery and invocation, following established practices in agentic systems and prior work on runtime code generation~\cite{fahim2026selfevolve}. Although in principle other models could be compared, we selected only one for the following reasons: 1) It is one of the few LLMs  with robust native function/tool-calling support, which is fundamental for matching user requests to functions (i.e., one of the main duties of the Chat \& Tool Dispatcher components), and it has been extensively used in software engineering experiments, particularly for code generation and agentic tasks~\cite{jiang2026survey,zhang2026survey}; 2) We are not concerned with the performance of the underlying LLM itself, but rather on the effect of the \textit{architectural} components of the pipeline. Model and architecture could in principle show interaction effects, but analysing this aspect would substantially increase the number of experiments required to achieve statistically valid results, and it is left for future work.
RE components use temperature $= 0.3$ for consistency; code generation uses temperature $= 0.7$ to explore diverse implementations. The maximum iteration count across both adjudicators is $K = 6$. Parameter values were determined through preliminary tuning experiments.

\textbf{Test-Code Collusion Mitigation.} Using a single model for both test and code generation creates collusion risk. Following established practices~\cite{huang2024agentcoder,dong2024selfcollaboration,fahim2026selfevolve}, we mitigate through: (1) role-separated agents with distinct system prompts, (2) fresh LLM API calls preventing shared context bias,
(3) adjudicator meta-validation improving test quality across iterations,
and (4) evaluation on novel problems not present in LLM training data.

\begin{table}[!t]
\centering
\setlength{\tabcolsep}{0.7pt}
\caption{Evaluation  Results.
}
\label{tab:results}
\footnotesize
\begin{tabular}{lcccc}
\toprule
\textbf{Project} & \textbf{LOC} & \textbf{ReqEvolve} & \textbf{SpecFix} & \textbf{Ablation} \\
\midrule
\textbf{Integration Tasks} & & & & \\
\hspace{3pt}Salary Analyzer & 603 & 87.5\% (14/16) & 37.5\% (6/16) & 31.2\% (5/16) \\
\hspace{3pt}Patient Risk Analyzer & 740 & 81.2\% (13/16) & 50.0\% (8/16) & 31.2\% (5/16) \\
\hspace{3pt}Student GPA Calculator & 773 & 93.8\% (15/16) & 93.8\% (15/16) & 50.0\% (8/16) \\
\hspace{3pt}Inventory Low Stock & 569 & 100.0\% (16/16) & 100.0\% (16/16) & 75.0\% (12/16) \\
\hspace{3pt}Healthcare Claims & 1097 & 100.0\% (16/16) & 100.0\% (16/16) & 81.2\% (13/16) \\
\hspace{3pt}Supply Chain Tracking & 3184 & 100.0\% (16/16) & 93.8\% (15/16) & 43.8\% (7/16) \\
\hspace{3pt}Student Analytics & 2041 & 93.8\% (15/16) & 87.5\% (14/16) & 81.2\% (13/16) \\
\hspace{3pt}E-Commerce Order & 2374 & 93.8\% (15/16) & 81.2\% (13/16) & 50.0\% (8/16) \\
\hspace{3pt}\textit{Subtotal} & & \textit{93.8\% (120/128)} & \textit{80.5\% (103/128)} & \textit{55.5\% (71/128)} \\
\midrule
\textbf{Compositional Tasks} & & & & \\
\hspace{3pt}Matrix Eigenvalue & 50-60 & 68.8\% (11/16) & 25.0\% (4/16) & 56.2\% (9/16) \\
\hspace{3pt}Portfolio Risk Calculator & 50-60 & 75.0\% (12/16) & 31.2\% (5/16) & 68.8\% (11/16) \\
\hspace{3pt}IoT Sensor Pipeline & 50-60 & 81.2\% (13/16) & 62.5\% (10/16) & 68.8\% (11/16) \\
\hspace{3pt}\textit{Subtotal} & & \textit{75.0\% (36/48)} & \textit{39.6\% (19/48)} & \textit{64.6\% (31/48)} \\
\midrule
\textbf{Data Processing Tasks} & & & & \\
\hspace{3pt}Movie API Interface & 105 & 87.5\% (14/16) & 100.0\% (16/16) & 43.8\% (7/16) \\
\hspace{3pt}Book Recommender & 105 & 87.5\% (14/16) & 62.5\% (10/16) & 50.0\% (8/16) \\
\hspace{3pt}Performance Tracker & 55 & 93.8\% (15/16) & 37.5\% (6/16) & 37.5\% (6/16) \\
\hspace{3pt}Friend Suggester & 2903 & 87.5\% (14/16) & 81.2\% (13/16) & 31.2\% (5/16) \\
\hspace{3pt}Sales Analytics & 2815 & 87.5\% (14/16) & 43.8\% (7/16) & 62.5\% (10/16) \\
\hspace{3pt}Server Log Analysis & 2969 & 93.8\% (15/16) & 87.5\% (14/16) & 62.5\% (10/16) \\
\hspace{3pt}Customer Analytics & 3124 & 93.8\% (15/16) & 93.8\% (15/16) & 93.8\% (15/16) \\
\hspace{3pt}\textit{Subtotal} & & \textit{90.2\% (101/112)} & \textit{72.3\% (81/112)} & \textit{54.5\% (61/112)} \\
\midrule
\textbf{Overall} & 23,622 & \textbf{89.2\% (257/288)} & 70.5\% (203/288) & 56.6\% (163/288) \\
\bottomrule
\end{tabular}
\end{table}

\subsection{Dataset}

We created 18 Python projects spanning healthcare,
finance, education, e-commerce, linear algebra, IoT, media systems, social networks, and HR analytics. Codebases range from 40 to 3,100 LOC. Although synthetic rather than industrial, the projects were designed to approximate realistic test settings covering different domains and codebase structures. Eleven projects extend the SelfEvolve codebases. Whereas in SelfEvolve the maximum codebase size was 783 LOC, our benchmark includes seven projects with over 2,000 LOC, indicating a substantial increase in structural complexity.
Table~\ref{tab:results} presents the complete project list.

We created 4 user stories per project to trigger self-evolution, designed with varying quality levels: 30.6\% of the user stories were injected with defects, and 69.4\% were crafted to be well-formed, based on the AQUSA quality criteria. The $\sim$70/30 split was chosen to approximate a realistic setting in which most user requests are usable, but a substantial minority still exhibit quality defects requiring clarification or repair. This allows the evaluation to reflect practical usage while still providing enough defective cases to meaningfully assess the robustness of the quality-assurance stage.

The projects, together with their user stories, span three self-evolution task categories. Integration tasks require generated functions to integrate with existing classes and methods in existing codebases. Data processing tasks require analysis functions for CSV/JSON datasets. Compositional tasks require reusing previously generated functions as building blocks for complex capabilities. This categorisation tests different generation challenges: codebase understanding and integration, data file handling and analysis, and function generation and reuse.

Overall, the benchmark comprises 18 projects and 72 user stories, representing 72 cases of self-evolution.

\subsection{Evaluation Methodology}

Evaluating user story-driven generation requires addressing a fundamental challenge: user stories permit multiple valid implementations. A request like ``discover people to connect with'' can be satisfied through mutual friends, shared interests, or geographic proximity, all of which satisfy the user's stated goal. Traditional assertion-based testing (e.g., \texttt{assert result == expected\_list}) assumes single correct outputs, incorrectly rejecting valid alternatives.
To address this issue, our evaluation employs an LLM-as-judge protocol. The judge evaluates each generated implementation against a set of predefined binary criteria, accepting or rejecting it based on its semantic alignment with the intended requirements. This enables implementation-level assessment without assuming a single canonical solution.
This is a limitation, since it does not fully capture the user perspective of validating functionality through observed behaviour. However, it complements rather than replaces execution-based checking, which is already enforced during generation through the TDD  pipeline.

Each binary criterion assesses whether a semantic requirement is met (e.g., ``Did it retrieve data?'') rather than prescribing implementation (e.g., ``Did it call get\_data()''?), enabling fair assessment across implementation diversity.
We adopt a binary forced-choice evaluation scheme to reduce inter-evaluator inconsistency. Empirical evidence shows that binary criteria demonstrate +45\% agreement improvement over Likert scales~\cite{checkeval2025,zheng2023judging}.

The authors defined 2-4 binary ground truth criteria per user story (number of criteria based on requirement complexity), before experiment execution,  following a systematic protocol: (1) \textit{Requirement extraction:} they analysed project codebase and brainstormed on possible valid outcomes that can satisfy each user story to identify essential operations (e.g., database retrieval, validation, filtering). (2) \textit{Question formulation:} converted requirements into binary Yes/No questions.

\textbf{LLM-as-Judge Reliability.} Our use of an LLM-as-judge protocol follows established practice for text-generation evaluation, notably MT-Bench~\cite{zheng2023judging} and criterion-based checklist frameworks~\cite{checkeval2025}, which have demonstrated substantial LLM-human agreement in analogous settings. To further address potential concerns about automated evaluation reliability, we implemented several safeguards. Each criterion is evaluated independently at temperature $= 0$ for deterministic judgment. The judge receives the user story $U$, generated code $F$, all classes and data files for each project, and acceptance criteria $AC$, enabling semantic reasoning rather than syntactic pattern matching. It verifies \textit{whether} a requirement is met, not \textit{how}, but needs to inspect the code to verify that a plausible output is not produced by an incorrect function. The final score is interpreted strictly: the function is \textbf{Correct} only when \textit{all} criteria pass; any failure scores as \textbf{Incorrect}.

To validate reliability, one researcher independently evaluated a stratified sample of $N = 267$ evaluation results drawn from 864 total runs at the 95\% confidence level with a 5\% margin. This manual evaluation work required approximately 16 hours. The sample was balanced across: (1) equal system and baselines representation (we will compare SelfEvolve with a similar baseline, and an ablation baseline, so we have 89 cases per system (267/3)), (2) proportional Pass/Fail distribution, and (3) coverage across all 18 projects. For each case, both the human rater and o4-mini LLM judge (a different model family from GPT-4.1) examined the user story, generated code, and all classes and data files for each project to determine whether all criteria are satisfied. Cohen's kappa measured inter-rater agreement at $\kappa = 0.67$, indicating substantial agreement~\cite{landis1977measurement}, with 82.8\% raw agreement (221/267 cases). Per-system kappa was consistent (\textsc{ReqEvolve} $\kappa = 0.62$, Ablation $\kappa = 0.58$, SpecFix $\kappa = 0.62$), confirming that agreement levels are stable across systems.

\textbf{Evaluation Metrics.} We measure \textbf{Pass@1}: the percentage of runs where generated code passes all binary criteria. Following standard code generation evaluation~\cite{humaneval}, Pass@1 indicates success in a single attempt. However, unlike traditional Pass@1, which checks functional correctness against test cases, our Pass@1 consider the binary criteria:  any single criterion failure scores the entire run as 0\%. Each user story is tested with 4 independent runs; we report average Pass@1 computed as the mean success rate across these 4 runs for each case, then averaged across all cases.

\subsection{RQ1: \textsc{ReqEvolve} Performance}

This question evaluates \textsc{ReqEvolve} performance based on the provided dataset, and considering the LLM-as-a-judge approach. Furthermore, we compare against SpecFix~\cite{specfix2025}, selected as the baseline. SpecFix is the state-of-the-art RE-focused system, demonstrating best performance among requirement engineering approaches as described in Section~\ref{sec:background}. SpecFix is also the conceptually closest baseline, as other code generation systems (MetaGPT, AgentCoder, AutoGen) focus on multi-agent coordination or planning-based generation rather than RE.
It should be remarked that SpecFix mostly targets \textit{technical} requirements, so the evaluation is admittedly not entirely fair, as the addressed problem is different.  However, this comparison remains valuable because it not only provides the strongest available RE-centered reference point, but also helps illustrate that user-driven runtime evolution is a distinct RE problem whose characteristics are not fully addressed by approaches developed for conventional technical requirements.

We compare \textsc{ReqEvolve} against SpecFix across all 72 user stories (18 projects $\times$ 4 story variants) with 4 independent runs each (288 evaluations per system). Both systems were evaluated using identical binary criteria on the same benchmark.
Table~\ref{tab:results} reports the results.
\textsc{ReqEvolve} achieves 89.2\% Pass@1 (257/288 successful runs), compared to SpecFix's 70.5\% (203/288), representing an 18.8\% improvement. Table~\ref{tab:results} presents complete results across all 18 projects.
Statistical significance was assessed via Wilcoxon signed-rank test on 18 paired projects. Projects serve as the independent unit of analysis, with performance aggregated across the 4 user stories per project, accounting for within-project dependencies. Analysis yields $W = 99.5$, $p < 0.01$, large effect $r = 0.79$, confirming statistically significant improvements.

Performance analysis by task type shows that compositional tasks are the ones where the difference between the approaches is more evident  ($35.4$\% difference), indicating that the ability to integrate previously generated functions is a distinguishing feature of \textsc{ReqEvolve}. On the other hand, these tasks are also the most challenging for \textsc{ReqEvolve} itself. This points to the need to inject a better understanding of the evolving codebase and improved architectural reasoning. Additional components (e.g., performing change impact analysis, storing an evolving architectural model) could improve this aspect.

\subsection{RQ2: Ablation Study}
This ablation study isolates automatic RE's contribution by comparing \textsc{ReqEvolve} against a baseline without RE components. The baseline replicates SelfEvolve, the only existing runtime self-extension framework, which shares identical infrastructure but omits automatic RE. Prior work~\cite{fahim2026selfevolve} evaluated SelfEvolve against alternative code-generation baselines such as AgentCoder, AutoGen, and MetaGPT, reporting superior performance.

Both systems were again evaluated on the same 72 user stories, 4 runs per story (288 runs per system, 576 total). The TDD component was already ablated by SelfEvolve, demonstrating substantial impact. The RE sub-components are tightly linked with each other, so they are not evaluated independently. The current evaluation comprises 864 runs across three systems; adding fine-grained ablations would increase this to over 1,000 runs, increasing resource use without substantial additional insight.

The comparison is presented in Table~\ref{tab:results}. \textsc{ReqEvolve} achieves 89.2\% Pass@1 (257/288), a 32.6\% improvement over the ablation baseline (56.6\%, 163/288). The majority of projects show substantial improvements, demonstrating that automatic RE benefits diverse task types. Integration and data processing tasks show larger improvements. These results suggest that requirement decomposition provides substantial value when generating code that integrates with existing components or the need to analyse data, hypothetically because decomposed requirements help the agent better navigate existing code and data slices. Compositional tasks (i.e., reusing generated functions) show smaller gains, suggesting that the TDD component that characterises SelfEvolve might be the core element enabling composition, especially considering the weak performance of SpecFix on these tasks, compared to the others, where it always outperforms SelfEvolve. Improvement of the TDD pipeline could thus also help achieve better performance in compositional tasks.

Wilcoxon signed-rank test on 18 paired projects yields $W = 153.0$, $p < 0.001$, large effect $r = 0.88$, confirming that automatic RE provides substantial benefit.

\subsection{RQ3: Failure Analysis}

To understand why \textsc{ReqEvolve} runs fail, we classified each of the 31 failed runs (10.8\% of 288) using the error$\to$fault$\to$failure chain of the development-perspective terminology standardized in ISO/IEC/IEEE 24765~\cite{iso24765}\footnote{IEEE standards (e.g., IEEE 1044-2009, ISO/IEC/IEEE 24765:2017) frame `error' as a human action; in our case it is the developer.}: an \textit{error} is an action that produces an incorrect result, a \textit{fault} is its manifestation in the software, and a \textit{failure} is the resulting inability to perform the required function. In our setting, the generation pipeline plays the developer's role, so errors are incorrect pipeline actions. We adopted qualitative coding rather than traditional root-cause analysis~\cite{ikram2022root,budhathoki2022causal}, as the latter assumes deterministic causal structures unsuited to probabilistic LLM generation~\cite{zhang2025failure}. Working backwards from observable outcome to root cause, we employed a systematic three-step procedure. First, we inspected the evaluation results to identify which criteria failed (the observable \textit{failure}). Second, we manually examined the generated code against the failing criteria to identify the specific code defect--- such as missing function calls, incorrect imports, or  flawed logic--- that caused the criterion to fail (the \textit{fault}). Third, for each fault, we inspected system logs, LLM prompts, and context retrieval outputs to hypothesise which pipeline condition produced it (the \textit{error}). For each case, the first author recorded structured notes documenting the chain. After annotating all 31 cases, we derived codes inductively by grouping similar patterns, iteratively refining labels until reaching a stable taxonomy. The process yielded 14  error$\to$fault$\to$failure chains.

The taxonomy (Table~\ref{tab:taxonomy}) uses four error codes (E), eight
fault codes (Fo), and two failure codes (F). Notably, no failures involved
code from unrelated problems contaminating the LLM input, nor did the model
generate code for a wrong problem entirely---confirming that the RE layer
correctly isolates and scopes each generation task.

\begin{table}[t]
\centering
\caption{Error-Fault-Failure Taxonomy and Frequencies, n=31}
\label{tab:taxonomy}
\footnotesize
\begin{tabular}{llr}
\toprule
\textbf{Code} & \textbf{Description} & \textbf{n (\%)} \\
\midrule
\multicolumn{3}{@{}l@{}}{\textit{Errors (Pipeline Conditions)}} \\
E1 & Wrong project files selected by LLM & 3 (9.7\%) \\
E2 & LLM produces incorrect code despite correct inputs & 20 (64.5\%) \\
E3 & Needed project files not selected by LLM & 2 (6.5\%) \\
E4 & LLM judge incorrectly scores working code as broken & 6 (19.4\%) \\
\multicolumn{3}{@{}l@{}}{\textit{Faults (Code Defects)}} \\
Fo1 & Code is only import statements and empty functions & 3 (9.7\%) \\
Fo2 & Code omits required operations & 5 (16.1\%) \\
Fo3 & Code does not use project's existing classes & 7 (22.6\%) \\
Fo4 & Code does not call previously generated functions & 7 (22.6\%) \\
Fo5 & Function name or parameters are incorrect & 2 (6.5\%) \\
Fo6 & Code attempts to call nonexistent functions & 3 (9.7\%) \\
Fo7 & Code calculates wrong metric & 2 (6.5\%) \\
Fo8 & Code stops after initial steps, not completing all & 2 (6.5\%) \\
\multicolumn{3}{@{}l@{}}{\textit{Failures (Evaluation Outcomes)}} \\
F1 & Code does not pass all evaluation criteria & 26 (83.9\%) \\
F2 & Code passes but LLM judge scores it as failing & 5 (16.1\%) \\
\bottomrule
\end{tabular}
\end{table}

\textbf{Errors.} E2 (incorrect code despite correct inputs) dominates with 20 cases (64.5\%), pointing to a limitation in LLM code generation capability even with well-formed inputs. E4 (judge errors, 6 cases, 19.4\%) occurs when the LLM judge evaluates functionally correct code as failing, suggesting a difficulty of LLM-based evaluation in assessing unconventional but valid implementations. Context-related errors E1 (wrong files selected) and E3 (needed files omitted) total 5 cases (16.1\%), indicating that the retrieval mechanism surfaces relevant files in the vast majority of runs; targeted improvements to file selection could further reduce this minority.

\textbf{Faults.} Integration-related defects dominate: Fo3 (does not use existing classes) and Fo4 (does not call previously generated functions) together account for 14 of 31 failures (45.2\%), particularly in existing compositional projects where multiple user requests build upon each other, requiring current generation to invoke functions from earlier requests. Completeness defects form the second cluster: Fo2 (omits required operations, 5 cases), Fo1 (empty scaffolding, 3 cases), and Fo8 (abandons generation after initial steps, 2 cases) together account for 10 cases (32.3\%), where the model produces partial solutions that address some but not all required functionality. Function call defects - Fo6 (calls nonexistent functions, 3 cases) and Fo5 (incorrect function names or parameters, 2 cases) - contribute 5 cases (16.1\%), while Fo7 (computes wrong metric, 2 cases, 6.5\%) represents  logic defects. The dominant chain E2$\to$Fo4$\to$F1 (7 cases) illustrates this: despite  producing functionally valid logic, the model ignores the project's existing codebase. Across all categories, the model struggles less with understanding \textit{what} to implement than with \textit{how} to fully integrate with and complete the task within existing structures---making codebase integration and task completeness explicit requirements could address this gap.

\textbf{Failures.} F1 (criteria not passed) accounts for 26 cases (83.9\%), while F2 (code passes but judge scores failing) accounts for 5 (16.1\%). These F2 cases originate in the LLM-as-judge component of the evaluation pipeline, not in code generation, and are not counted as generation failures. Of all failures, 25 (80.6\%) pass at least one criterion, indicating that \textsc{ReqEvolve} produces partially correct output even when failing.  Excluding the 5 F2 cases attributable to the LLM-as-judge, the generation failure rate is 9.0\% (26/288).

It should be noted that \textsc{ReqEvolve} achieves higher Pass@1 on defective user stories (93.2\%) than well-formed stories (87.5\%).  Most defective stories in our benchmark are under-specified: for example, they omit the role or benefit. Under-specified stories can admit a wider set of acceptable implementations, while well-formed stories constrain the solution space more tightly. Thus, the better result obtained with defective user stories likely reflects a broader semantic acceptance space together with a successful clarification within that space; not all implementations are acceptable, just more than those that would satisfy a clearer request. This is in line with the final target of enabling users to incrementally converge towards a satisfactory implementation after seeing a prototype of the change, starting from vaguely expressed ideas, as it happens in practice. This does not mean that all defect types lead to successful implementations: a systematic variation of the effects of different defect types, as done in~\cite{vogelsang2025impact}, is left for future work. We refrain from statistical testing as user stories within projects are not fully independent samples.

\section{Discussion and Future Plans}

\textbf{Threats to Validity.} \textit{Internal Validity.} (1) Our manual review of a stratified sample of evaluation results was performed by a single reviewer, introducing potential subjectivity. However, the structured binary criteria reduce subjective judgment compared to open-ended assessment, and LLM-as-a-judge has already been confirmed as a viable solution for evaluation~\cite{zheng2023judging}. (2) Using GPT-4.1 for both test and code generation creates collusion risk. This is mitigated through architectural separation (cf.\ Sect.~4.1.3), following established practices~\cite{huang2024agentcoder,dong2024selfcollaboration}. (3) Generated functions may depend on previously-generated or pre-existing   functions, meaning capabilities are not fully independent. We do not perform regression testing; a new function could break existing ones through shared state or dependency chains.

\textit{External Validity.} (1) The benchmark spans 18 Python projects across diverse domains, but emerging domains remain unexplored and results may not generalize to all software contexts and programming languages. It should be noted that our current benchmark evaluates the technical viability of user-driven self-evolution in a controlled setting, focusing on proof of concept and core component validation. A real-world study of the paradigm would require a different design, using real-world codebases and end users for request formulation; our own limited domain and code knowledge would restrict our ability to meaningfully manipulate queries and tasks in such a setting. (2) Safety-critical domains such as medical device control and aviation software require stronger verification processes that our binary criteria evaluation cannot provide. These types of contexts are out of scope, and our idea focuses on consumer software and non-mission-critical solutions. (3) We employed only GPT-4.1, selected because our architecture relies on its function-calling capability for discovering and executing existing functions. Our evaluation focuses on cases where generation is needed; we do not evaluate the function-calling component itself, as this relies on OpenAI's capability rather than our contribution. Furthermore, GPT-4.1 is by now a relatively older model, and model-specific findings are inherently prone to obsolescence as newer models appear; whether the reported gains persist under more recent LLMs is left as future work. (4)  User story complexity is bounded by single-function scope: transfer is expected mainly for incremental, additive requests where the generated capability can be implemented as a localized function reusing existing code without disrupting the architecture. This matches the current scope of self-extension, rather than self-refactoring. Performance is expected to decrease for industrial requirements involving complex cross-cutting changes or deeper refactoring. (5) Our evaluation assesses functional correctness only, since user stories inherently focus on functional capabilities; non-functional aspects and code quality dimensions are outside our scope. (6) While our system is designed to enable multiple user interaction and clarification cycles, at this stage, we have evaluated it considering a ground truth to obtain a controlled and repeatable evaluation, rather than an interactive user session. This is a necessary first step: an interactive user study is meaningful only after showing that the pipeline can generate sufficiently accurate capabilities in a controlled setting. Less constrained user studies are left as future work.

\textit{Construct Validity.} (1) The binary Pass@1 metric aggregates all-or-nothing success, potentially obscuring cases where implementations satisfy most but not all criteria. Alternative metrics (average criterion satisfaction, weighted criteria) would reveal gradations but sacrifice the clear success/failure boundary that binary evaluation provides. (2) Evaluation relies on LLM-as-judge for binary criterion assessment. While LLM judges can exhibit systematic biases, our manual validation of a stratified sample provides empirical support for reliability in this context, and the structured binary format reduces ambiguity compared to open-ended evaluation. (3) User stories in the benchmark contain quality defects, as demonstrated by the AQUSA validation results and the Clarification component's activation on defective stories. This validates the components designed to detect and address such defects. The remaining threat is whether the defect distribution in our benchmark represents real-world defect distributions; more systematic evaluation across defect types is needed. (4) Our predefined ground truth criteria may not cover all valid implementations. Therefore, our results represent a lower bound of acceptable solutions.

\textbf{Discussion and Future Work.} \textit{Implications for RE  Research.} Most LLM-for-RE research focuses on isolated RE tasks such as requirements classification or traceability link recovery~\cite{vogelsang2025impact}. \textsc{ReqEvolve} takes a step forward by integrating RE into a complete software engineering pipeline, combining quality assessment, clarification, specification decomposition, and code generation within a single runtime process. This positions RE not as a standalone activity but as an embedded component of LLM-based SE pipelines, and further studies should investigate the effect of RE pipelines in other complex tasks such as model generation, test generation, change impact analysis, and refactoring. This integration also redefines the role of requirements analysts: rather than writing specifications manually, analysts shift toward curating RE components, defining quality thresholds, and designing clarification strategies that guide automated pipelines.

A broader implication is that user-driven self-evolution enables fully individualised software: users can  start from a \textit{minimal} implementation and add capabilities upon request. Indeed, the continuous growth of application size, driven by unrequested features and updates, contributes to making devices obsolete, and personalised software that includes only the features actually needed by each user offers an alternative. In addition, the two-tier validation model (user validation for experimental use, developer certification for diffusion) suggests new human-in-the-loop paradigms for requirements validation where users validate intent match while developers certify technical correctness, creating novel division-of-labor patterns in requirements processes. This impacts RE research by demonstrating that requirements validation can occur through user interaction with working implementations rather than solely through documentation review. This suggests that RE processes could incorporate rapid prototyping earlier in the validation cycle, enabling users to validate intent through observable behavior rather than abstract specifications.

\textit{Implications for Code Generation Research.} The substantial improvements over both SpecFix (RQ1) and the ablation baseline (RQ2) demonstrate that RE is not merely beneficial but essential for user story-driven generation.
Our results suggest that augmenting code generation architectures with structured requirement processing provides substantial gains, motivating research into hybrid architectures combining LLM reasoning with structured RE transformations. The binary criteria evaluation framework addresses a methodological gap in code generation benchmarking: existing benchmarks (HumanEval, MBPP) assume single correct implementations, limiting the evaluation of open-ended generation tasks. Our framework of semantic requirement questions with LLM-based assessment generalises to other domains requiring implementation-agnostic correctness evaluation. The manual validation of a stratified sample of 267 LLM judgments provides empirical support for LLM-as-judge reliability in code evaluation contexts, extending prior work on NLP task evaluation~\cite{checkeval2025} to code generation.

\textit{Implications for Software Engineering Practice.} The individualised software paradigm enabled by user-driven self-evolution creates novel software process considerations. Traditional processes assume developers mediate all capability changes; runtime self-evolution introduces direct user-to-system capability requests.
The two-tier validation model provides a starting framework but requires empirical validation through deployment studies. For organizations adopting this model, we recommend: (1) sandboxed experimental zones where users can freely generate capabilities for personal use without affecting shared infrastructure, (2) review boards examining generated capability logs before organizational diffusion, (3) usage monitoring identifying frequently-requested capabilities worthy of developer-certified implementation.

\textit{Future Directions.}
(1) \textit{Multi-modal user input} could extend beyond text-based user stories to accept voice descriptions, annotated screenshots, or GUI demonstrations, broadening accessibility for non-technical users. (2) \textit{User studies} can investigate whether end users are able to express capability needs effectively through user stories in practice, how much clarification is typically required, and whether interaction with generated behaviour improves requirements validation compared to conventional elicitation and prototyping workflows. Such studies should focus on user-relevant metrics (e.g., user satisfaction) and qualitative observations of how users reformulate requests after observing the generated output, rather than only pipeline correctness. Iterative refinement is expected to help when users converge on incremental requests after seeing a concrete implementation, since prototypes can reveal needs that were initially implicit; interaction may also introduce requirement drift, changing expectations, or other phenomena not captured by our current benchmark.
(3) \textit{Multi-function coordination} generating coordinated function sets from epic-level stories, handling interdependencies and shared state across multiple generated capabilities. (4) \textit{Additional RE components} can be introduced in the pipeline, for example, to support change impact analysis and refactoring, to deal with larger and more realistic projects. (5) \textit{Proactive self-evolution via MAPE-K}~\cite{mape-k} extending from reactive (user-triggered) to proactive generation, where the system monitors function usage patterns and self-initiates improvements~\cite{li2024genaisas}. (6) \textit{Cross-model replication} can assess whether the reported gains transfer to more recent function-calling-capable models beyond GPT-4.1, isolating the effect of the RE pipeline from the underlying LLM. (7) \textit{Real-world function-level benchmarks} such as CoderEval~\cite{yu2024codereval} are considered as a next step, since their function-level tasks better match our current single-function self-extension scope, and we expect similar transfer trends under this adaptation. This requires defining user-level requests corresponding to the target functions, since CoderEval itself is not formulated as user-driven self-evolution.

\textbf{Conclusion.} We presented \textsc{ReqEvolve}, a runtime code generation system for user-driven self-evolution that accepts user stories as input and integrates automatic RE into the generation pipeline. Across 72 software evolution cases spanning 18 projects, the system achieved 89.2\% Pass@1, outperforming SpecFix by 18.8\%, and the ablation baseline by 32.6\%. These results show that structured RE support is not peripheral but foundational for user story-driven runtime generation. More broadly, the work points toward a software process in which requirements validation can occur through immediate interaction with generated behaviour, while technical verification remains a subsequent developer responsibility.

\noindent
\textbf{Data Availability.} Replication Package at~\cite{fahim2026replication}.

\noindent
\textbf{AI Disclosure:} We used ChatGPT 5.1 for revision.

\bibliographystyle{ACM-Reference-Format}
\bibliography{references}

\end{document}